\documentclass[aps,prb,twocolumn,groupedaddress,showpacs]{revtex4}

\usepackage{graphicx}

\begin{document}

\title{Crossover in the nature of the metallic phases in the perovskite-type $R$NiO$_3$}

\author{K.~Okazaki$^1$, T.~Mizokawa$^1$, A.~Fujimori$^1$,
E.V.~Sampathkumaran$^2$, M.J.~Martinez-Lope$^3$ and
J.A.~Alonso$^3$}
\affiliation{$^1$Department of Physics and Department of
Complexity Science and Engineering, University of Tokyo,
Bunkyo-ku, Tokyo 113-0033, Japan} 
\affiliation{$^2$Department of Condensed Matter Physics and Materials
Science, Tata Institute of Fundamental Research, Colaba, Mumbai 400-005,
India} 
\affiliation{$^3$Instituto de Ciencia de Materiales de Madrid,
CSIC, Cantoblanco, E-28049 Madrid, Spain}

\date{\today}

\begin{abstract}
 We have measured the photoemission spectra of
 Nd$_{1-x}$Sm$_{x}$NiO$_{3}$, where the metal-insulator transition and
 the N\'{e}el ordering occur at the same temperature for $x \lesssim
 0.4$ and the metal-insulator transition temperature ($T_{MI}$) is
 higher than the N\'{e}el temperature for $x \gtrsim 0.4$. For $x \le
 0.4$, the spectral intensity at the Fermi level is high in the metallic
 phase above $T_{MI}$ and gradually decreases with cooling in the
 insulating phase below $T_{MI}$ while for $x > 0.4$ it shows a
 pseudogap-like behavior above $T_{MI}$ and further diminishes below
 $T_{MI}$. The results clearly establish that there is a sharp change in
 the nature of the electronic correlations in the middle ($x \sim 0.4$)
 of the metallic phase of the $R$NiO$_3$ system.
\end{abstract}


\maketitle

\begin{figure}[b]
\includegraphics[width=8.5cm]{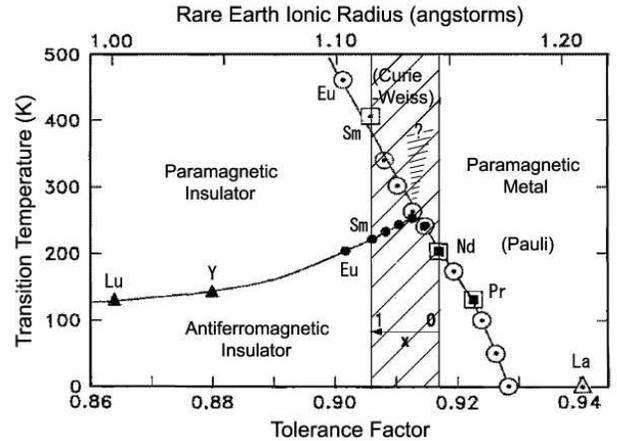} 
\caption{\label{fig1}Phase diagram of $R$NiO$_3$ taken from
 Ref.~\onlinecite{phase}. The hatched area indicates the composition
 range for which we have measured the photoemission spectra and the
 dashed lines indicates a new crossover or phase boundary line in the
 metallic phase proposed in this study.}
\end{figure}

Many $3d$ transition-metal compounds undergo a metal-insulator
transition as a function of temperature, carrier concentration, band
width and so on.~\cite{Mott,MIT} In these systems, electron correlation
plays an important role. V$_2$O$_3$, NiS$_{2-x}$Se$_x$, NiS and
$R$NiO$_3$ ($R$: rare earth) are well-known {\it bandwidth-controlled}
metal-insulator transition systems. They undergo a transition from an
antiferromagnetic insulator to a paramagnetic metal under high pressure
or under ^^ ^^ chemical pressure''. To understand their phase diagrams
from a microscopic point of view is one of the major goals in the
studies of strongly correlated systems and much theoretical effort has
been made so far along this direction.~\cite{DMFT,Takano} In V$_2$O$_3$
(Ref.~\onlinecite{V2O3}) and NiS$_{2-x}$Se$_x$
(Ref.~\onlinecite{NiSSe}), the metallic phase exists on the lower
temperature side of the paramagnetic insulating phase as predicted
theoretically~\cite{DMFT} and as expected from the large magnetic
entropy in the paramagnetic insulating phase. On the other hand, the
phase diagram of $R$NiO$_3$ has a peculiar feature that the metallic
phase exists on the higher temperature side of the paramagnetic
insulating phase as shown in Fig.~\ref{fig1}.~\cite{phase}

The perovskite-type oxide $R$NiO$_3$ belongs to the charge-transfer
regime of the Zaanen-Sawatzky-Allen (ZSA) diagram\cite{ZSA}, and its
ligand-to-metal charge-transfer energy $\Delta$ is smaller than the
$3d$-$3d$ Coulomb repulsion energy $U$ ($\Delta < U$). Hence its band
gap is given by $\sim \Delta-W$, where $W$ is the band width. $W$
depends on the Ni-O-Ni bond angle or the tolerance factor [$\equiv
d_{R{\rm -O}}/\sqrt{2}d_{\rm Ni-O}$], which depends on the rare-earth
ionic radius. Hence, whether the band gap is open ($\Delta \gtrsim W$)
or closed ($\Delta \lesssim W$) is determined by the deviation of the
Ni-O-Ni bond angle from the ideal value 180$^\circ$. As illustrated in
Fig.~\ref{fig1}, the least distorted LaNiO$_3$, which has the perovskite
structure with a weak rhombohedral distortion, is a paramagnetic metal
at all temperatures. The other $R$NiO$_3$ members, which have the
orthorhombic GdFeO$_3$-type structure, undergo a temperature-induced
metal-insulator transition. The metal-insulator transition and the
antiferromagnetic ordering occur at the same temperature for $R$ = Pr
and Nd, while the metal-insulator transition temperature ($T_{MI}$) is
higher than the N\'{e}el temperature ($T_N$) for the compounds with $R$
= Sm and other small rare earths. As a result, the metallic phase exists
on the higher temperature side of the paramagnetic insulating phase. The
boundary between these two types of phase transitions is located between
NdNiO$_3$ and SmNiO$_3$.

Recently, Vobornik {\it et al.}~\cite{Vobornik} reported that the
temperature dependence of photoemission spectra near $E_F$ in the
insulating phase is different between NdNiO$_3$ and SmNiO$_3$, which is
probably related with the different magnetoresistance behaviors reported
by Mallik {\it et al}.~\cite{Mallik} In the Nd$_{1-x}$Sm$_{x}$NiO$_3$
system, $T_{MI} > T_N$ for $x \gtrsim 0.4$, while $T_{MI} = T_N$ for $x
\lesssim 0.4$ (Fig.~\ref{fig1}). In this Rapid Communicattion, we report
on a systematic photoemission study of Nd$_{1-x}$Sm$_x$NiO$_3$ as a
function of $x$ and temperature and reveal quite different behaviors
between $x \le 0.4$ and $x > 0.4$. Based on this result, we propose that
there is a phase boundary or a crossover line within the metallic phase,
which separates two different kinds of metallic phases in the $R$NiO$_3$
system.

\begin{figure}[t]
\includegraphics[width=8.5cm]{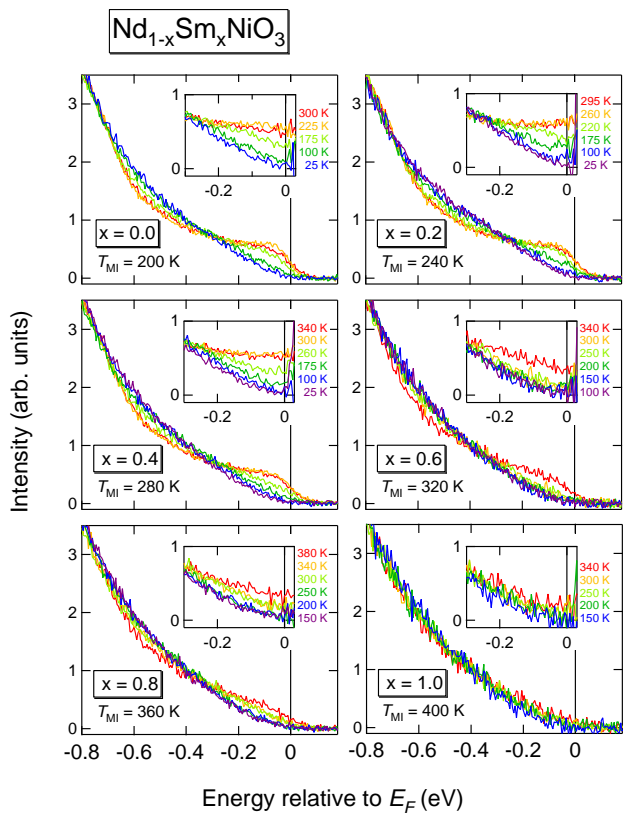}
\caption{\label{fig2}Temperature dependence of photoemission
spectra in Nd$_{1-x}$Sm$_x$NiO$_3$ for various $x$'s. Inset for
each panel shows spectra divided by the Fermi-Dirac function.}
\end{figure}
Preparation and characterization of polycrystalline
Nd$_{1-x}$Sm$_x$NiO$_3$ ($x$ = 0.0, 0.2, 0.4, 0.6, 0.8, 1.0) are
described elsewhere.~\cite{Mallik} The $T_{MI}$ was measured by the
differential scanning calorimetric method, and was 199.5 K for NdNiO$_3$
and 400.2 K for SmNiO$_3$. For the $T_{MI}$ of the intermediate
compounds, we have linearly interpolated between NdNiO$_3$ and SmNiO$_3$
according to Ref.~\onlinecite{Frand}. Photoemission measurements were
carried out using a VSW hemispherical analyzer and a VG He discharge
lamp. The He {\footnotesize I} (21.2 eV) resonance line was used for
excitation. The total energy resolution was set to about 30 meV.  Clean
surfaces were obtained by repeated {\it in situ} scraping at each
measurement temperature. The base pressure of the spectrometer was
better than 1 $\times$ 10$^{-10}$ Torr.

\begin{figure}[t]
\includegraphics[width=8cm]{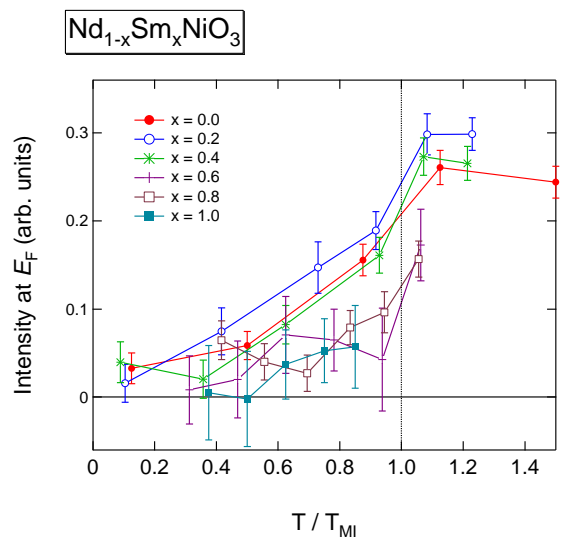}
\caption{\label{fig3}Photoemission intensity at $E_F$ as a
function of normalized temperature $T/T_{MI}$.}
\end{figure}
Figure~\ref{fig2} shows the photoemission spectra of
Nd$_{1-x}$Sm$_x$NiO$_3$ at various temperatures. All the spectra have
been normalized to the integrated intensity in the indicated energy
range, that is, from - 0.8 eV to 0.2 eV. The results for NdNiO$_3$ and
SmNiO$_3$ are consistent with Vobornik {\it et al.}'s
results,~\cite{Vobornik} i.e., the spectra of NdNiO$_3$ show a high
intensity at $E_F$ above $T_{MI}$ and decreasing intensity below
$T_{MI}$. The intensity continuously decreases with decreasing
temperature even well below $T_{MI}$. The spectra of SmNiO$_3$ show a
weaker temperature dependence below $T_{MI}$. As for
Nd$_{1-x}$Sm$_x$NiO$_3$, the spectra for $x$ = 0.2 and 0.4 show a strong
temperature dependence similar to NdNiO$_3$. They show clear spectral
weight transfer from the region - (0-0.3) eV to - (0.3-0.6) eV in going
from above $T_{MI}$ to well below it. The spectra for $x$ = 0.6 and 0.8
also show similar spectral weight transfer but the overall intensity is
weak similar to SmNiO$_3$. In the $x$ = 0.6, 0.8, and 1.0 compounds, the
$T_N$ is different from the $T_{MI}$, and no remarkable change has been
observed at $T_N$. The temperature dependence of the intensity at $E_F$
plotted in Fig.~\ref{fig3} clearly indicates the contrasting behavior
between $x \le 0.4$ and $x > 0.4$. Since the boundary between $T_{MI} =
T_N$ and $T_{MI} > T_N$ is located at $x \sim 0.4$, the result implies
that the electronic structure is different depending on whether $T_{MI}
= T_N$ or $T_{MI} > T_N$.
\begin{figure}[t]
\includegraphics[width=6.5cm]{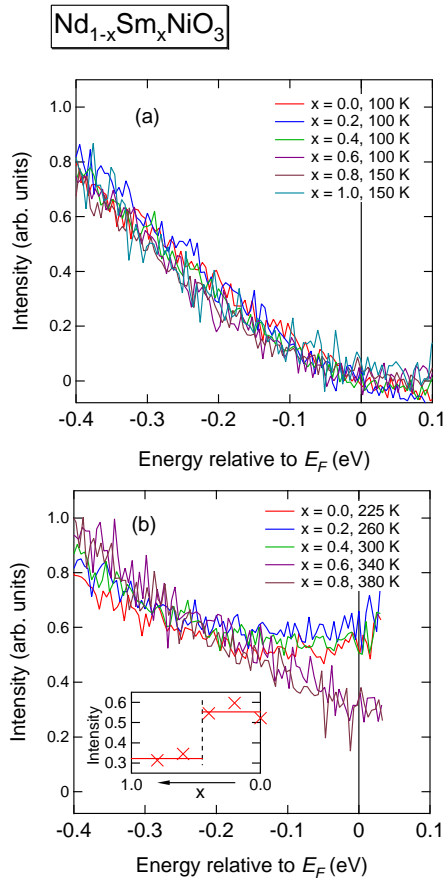}
\caption{\label{fig4}Composition dependence of the photoemission
spectra Nd$_{1-x}$Sm$_x$NiO$_3$ (a) Spectra taken at the lowest
temperatures. (b) Spectra taken just above $T_{MI}$. Inset shows
the intensity at $E_F$ just above $T_{MI}$ as a function of $x$.}
\end{figure}

In order to remove the effect of the Fermi-Dirac distribution function
and to deduce the experimental ^^ ^^ density of states" (DOS), we have
divided the spectra by the Fermi-Dirac distribution function broadened
with a Gaussian corresponding to the experimental resolution as shown in
the inset of each panel of Fig.~\ref{fig2}. For $x \le 0.4$, the
experimental DOS above $T_{MI}$ is almost flat or weakly increases with
energy. Just below $T_{MI}$, the DOS is reduced, but remains finite at
$E_F$, indicating a pseudogap opening of 0.1 - 0.2 eV. On the other
hand, the spectra for $x > 0.4$ is pseudogap-like already above
$T_{MI}$.

Figure~\ref{fig4} (a) and (b) shows the Fermi-Dirac function-divided
spectra taken at the lowest temperatures and just above $T_{MI}$,
respectively. The spectra taken at the lowest temperatures are identical
for all $x$'s. This suggests that the ground state is essentially the
same for all $x$'s, i.e., in the same antiferromagnetic insulating
state. This is in accordance with the fact that the magnetic structure
is the same between NdNiO$_3$ (Ref.~\onlinecite{neutron1}) and SmNiO$_3$
(Ref.~\onlinecite{neutron2}). Also, charge disproportionation was
recently found in NdNiO$_3$,~\cite{Zaghrioui} which implied that it is
present in all $R$NiO$_3$ ($R \ne$ La) because it had been already found
in other $R$NiO$_3$ with small $R$ ($R$ = Ho, Y, Er, Tm, Yb and
Lu).~\cite{Alonso} On the other hand, the spectra above $T_{MI}$ is
quite different between $x \le 0.4$ and $x > 0.4$. The spectra for $x
\le 0.4$ is nearly flat around $E_F$ whereas the spectra for $x$ = 0.6
and 0.8 show a weak pseudogap-like behavior. This indicates that the
paramagnetic metallic phase is different between $x \le 0.4$ and $x >
0.4$. As shown in the inset of Fig.~\ref{fig4} (b), the intensity at
$E_F$ just above $T_{MI}$ sharply changes at $x \sim 0.4$. From these
results, we consider that a phase boundary or a sharp crossover line
exists within the high-temperature metallic phase at $x \sim 0.4$.

There have been some reports which indicate that the metallic phase may
be different between NdNiO$_3$ and SmNiO$_3$. According to the magnetic
susceptibility measurement of NdNiO$_3$\cite{NdNiO3} and
SmNiO$_3$,~\cite{SmNiO3} the metallic phase of NdNiO$_3$ is
Pauli-paramagnetic after the subtraction of the rare-earth local-moment
contribution, while that of SmNiO$_3$ is Curie-Weiss like. This result
would indicate that the metallic phase of NdNiO$_3$ is essentially the
same as that of Pauli-paramagnetic LaNiO$_3$, while conduction electrons
in the metallic phase of SmNiO$_3$ show local moment behavior. Due to
the existence of the local moment, the metallic phase of SmNiO$_3$ would
have a large magnetic entropy. If the entropy of the local moments plus
that of the conduction electrons in the metallic phase exceed the
entropy of the local moments in the paramagnetic insulating phase, the
metallic phase can exist on the higher temperature side of the
paramagnetic insulating phase across the first-order metal-insulator
phase boundary. In the Nd$_{1-x}$Sm$_x$NiO$_3$ system, the band width
becomes narrower with Sm content, making $\Delta/W$ larger. We propose
that with Sm content, the Ni $3d$ electrons become more strongly
correlated and the local magnetic moment is induced.

Finally, we comment on the origin of the strong temperature dependence
of the spectra in the insulating phase. Granados {\it et
al.}\cite{Granados1,Granados2} and Blasco and Garcia\cite{Blasco} have
reported that hysteresis below $T_{MI}$ extends to a wide temperature
range up to 70 K from the transport and calorimetric measurement and
proposed that the metallic and insulating phases coexist over this
temperature range. The temperature dependence of the photoemission
spectra in the insulating phase may be related to this unusually strong
hysteresis. We may attribute these strong hysteretic features to
disorder such as oxygen non-stoichiometry and/or rare-earth
atom vacancies, which cannot be avoided in this kind of materials. In a
system where both disorder and electron correlation effect are
important, the spectra may show an unusual temperature dependence as
reported by Sarma {\it et al.}~\cite{Sarma} on disorder-induced
metal-insulator transition in LaNi$_{1-x}M_x$O$_3$ ($M$ = Mn, Fe, and
Co). According to them, while an insulating compound has a finite gap at
$E_F$ at very low temperatures, the gap closes at elevated
temperature. Competing or cooperative behaviors between electron
correlation and disorder have not been investigated so far
and have to be clarified in future.

In conclusion, we have studied the temperature-dependent
electronic structure of Nd$_{1-x}$Sm$_x$NiO$_3$ by photoemission
measurements. While the spectra at the lowest temperatures are
identical for all $x$'s, the spectra above $T_{MI}$ show a
pseudogap behavior for $x > 0.4$, different from the spectra of a
typical metal for $x \le 0.4$. It appears that the appearance of
the pseudogap is related with the appearance of local magnetic
moment in the metallic phase of $x > 0.4$. We propose that the
nature of the metallic state is different between these
two composition ranges caused by a change in the strength of
electron correlation between the two regions.

The authors would like to thank D. D. Sarma for enlightening
discussion. This work was supported by Grants-in-Aid for
Scientific Research A12304018 and ^^ ^^ Novel Quantum Phenomena in
Transition Metal Oxides'' from the Ministry of Education, Culture,
Sports, Science, and Technology.

\end{document}